\documentclass[12pt]{article}
\pdfoutput=1

\usepackage{geometry,amsmath,amsfonts}
\usepackage{slashed}
\usepackage{graphicx}
\usepackage{epstopdf}
\usepackage{mathrsfs}
\usepackage{amssymb}
\usepackage{verbatim}
\usepackage{color}
\usepackage{multirow}
\usepackage{subfig}
\usepackage{cite}

\begin{document}


\vspace*{8mm}

\begin{center}

{\large\bf  The Higgs-portal for vector Dark Matter and the Effective }\smallskip

{\large\bf Field Theory approach: a reappraisal}

\vspace*{9mm}

{\sc Giorgio Arcadi$^{1}$},  {\sc Abdelhak~Djouadi$^{2}$} and {\sc Marumi Kado$^{3,4}$}

\vspace*{9mm}

$^1$ Dipartimento di Matematica e Fisica and INFN, Universit\`a di Roma
Tre, Via della Vasca Navale 84, 00146, Roma, Italy. \\ \vspace{0.15cm}


$^2$ Universit\'e Savoie--Mont Blanc, USMB, CNRS, LAPTh, F-74000 Annecy, France. \\ \vspace{0.15cm}

$^3$ Department of Physics and INFN, ``Sapienza" Universt\`a di Roma, Pizzale Aldo Moro 5, I--00185 Roma, Italy.\\ \vspace{0.15cm}

$^4$ LAL, Universit\'e Paris-Sud, CNRS/IN2P3, Université Paris-Saclay, Orsay, France.\\ \vspace{0.15cm}


\end{center}

\vspace*{4mm}

\begin{abstract}
We reanalyze the effective field theory approach for the scenario in which the particles that account for the dark matter (DM) in the universe are vector states that interact only or mainly through the Standard Model-like Higgs boson observed at the LHC. This model-independent and simple approach, with a minimal set of new input parameters, is widely used as a benchmark in DM searches and studies in astroparticle and collider physics. We show that this effective theory could be the limiting case of ultraviolet complete models,
taking as an example the one based on a spontaneously broken U(1) gauge symmetry that incorporates a dark gauge boson and an additional scalar that mixes with the standard Higgs boson. Hence, despite the presence of the new degrees of freedom, measurements of the invisible decay branching ratio of the Higgs boson, as performed at colliders such as the CERN LHC, can be interpreted consistently in such an effective framework and can be made complementary to results of DM searches in direct detection experiments. 

\end{abstract}

\newpage

\section{Introduction}

Particle physics proposes a compelling solution to the puzzle of the missing or dark matter (DM) in the Universe, in terms of a colorless and  electrically neutral weakly interacting  massive particle (WIMP) that is stable at cosmological times and has a mass in the vicinity of the electroweak scale. Such WIMPs are predicted in many extensions of the Standard Model (SM); for reviews, see 
Refs.~\cite{Bertone:2004pz,Arcadi:2017kky} for instance. A very interesting class of such models is the one in which the DM states interact only through their couplings with the Higgs sector of the theory, the so--called Higgs--portal DM models; see Ref.~\cite{Arcadi:2019lka} for a recent review. The observed cosmological relic abundance would be  induced when pairs of DM states annihilate into SM fermions and gauge bosons, through the $s$--channel exchange of the Higgs bosons, and the latter will in turn be the mediators of the mechanisms that allow for the experimental detection of the DM states.

The simplest of the Higgs--portal scenarios is when the Higgs sector is minimal 
and, thus, identical to the SM one, namely a single doublet Higgs field structure that leads to the unique $H$ boson observed so far \cite{Aad:2012tfa,Chatrchyan:2012xdj}. One could then minimally extend the model by simply adding one new particle, the DM state, as an isosinglet under the electroweak group.  Nevertheless, the DM particle can have the three possible spin assignments and be a scalar, a vector or a Dirac or Majorana spin--$\frac12$ fermion. Although only effective, 
this approach can be adopted as it is rather model--independent and does not make any assumption on the very nature of the DM \cite{Kim:2006af,Kanemura:2010sh,Djouadi:2012zc,Djouadi:2011aa,LopezHonorez:2012kv,Goodman:2010ku,Fox:2011pm,Buckley:2014fba,Abdallah:2015ter,Alanne:2017oqj}.
Furthermore, it bears the advantage of having a very restricted number of extra parameters in addition to the SM ones \cite{Djouadi:2005gi}, namely the mass of the DM particle and its coupling to the Higgs boson. 

Such a minimal scheme has been investigated extensively  
and has been probed in direct and indirect detection in astrophysical experiments, 
and at colliders such as the CERN LHC. In the latter case, light DM particles are searched for by looking at invisible Higgs decays and other missing transverse energy signatures, making the two types of searches highly complementary. In particular, the  constraints from the LHC that can be derived on the mass of the DM particle and its couplings to the Higgs boson from the invisible Higgs decay rate, can be directly compared to the limits on the elastic  scattering cross section of the DM with nuclei, obtained in a direct detection experiment such as XENON1T \cite{Aprile:2017aty}. The ATLAS and CMS collaborations have 
interpreted their analyses of the invisible decays of the Higgs boson, in this framework 
since Refs.~\cite{Aad:2015pla,Chatrchyan:2014tja}. 

Nevertheless, this effective field theory (EFT) approach suffers from a drawback: except for the scalar DM case, the Higgs--portals are not complete models in the ultraviolet (UV) regime. UV--completeness that would make 
the Higgs--portal models more realistic, calls for additional degrees of freedom. In such a case, complementarity between astroparticle and collider constraints cannot be, in principle, established in a general and model--independent fashion and, thus, one needs to rely on the details of the specific UV--complete models.

This problem appeared to be particularly severe in the case of vectorial DM states. Early studies like the one conducted in Ref.~\cite{Lebedev:2011iq} have already shown that, because of perturbative unitarity bounds, the effective description becomes questionable at DM masses below a few tens of GeV. When later on, Refs.~\cite{Baek:2012se,Baek:2014jga} raised the issue that a renormalizable realization of the vector Higgs--portal scenario typically predicts the existence of new light degrees of freedom that make the correlation between the collider and direct search bounds less trivial, the description of the spin--1 vector DM case has been removed from the invisible Higgs analyses performed by the ATLAS and CMS collaborations~\cite{Aaboud:2019rtt,Aaboud:2018sfi,Sirunyan:2018owy,Khachatryan:2016whc}.   

The aim of the present work is to revisit the issue of a consistent interpretation of current and future experimental results in searches of invisible Higgs decays in the context of effective Higgs--portal models for a vectorial DM state.  We investigate whether the vector Higgs--portal case can be regarded as a consistent EFT limit of a UV--complete model.
If this is indeed the case, the correlation between constraints from invisible Higgs decays at the LHC and direct detection experiments such as XENON1T are valid. To do so, we compare the predictions of the EFT vector Higgs--portal scenario with those of one of its simplest completions in which the DM is identified with the stable gauge boson of a dark U(1) gauge group that is spontaneously broken by the vacuum expectation value of an additional complex scalar field \cite{Hambye:2008bq,Lebedev:2011iq,Baek:2012se,Farzan:2012hh,Arcadi:2016qoz,Glaus:2019itb}. 
The latter features mass mixing with the SM Higgs doublet that leads to the 125 GeV $H$ boson and the portal between the DM and the SM particle sectors is then represented by the two Higgs mass eigenstates. 

We show that in this model, despite of perturbative unitarity which constrains the mass hierarchy between the additional and the SM--like Higgs states, the EFT limit can be indeed approximately recovered. This occurs in the case where the new Higgs particle is sufficiently heavy and its mixing with the SM--like one is sufficiently suppressed.

The remainder of the paper is structured as follows. In the next section, we present the theoretical aspects of the effective Higgs--portal model and its ultraviolet
complete realization based on a spontaneously broken U(1) gauge  group. In section 3, we compare the outcome of the two approaches in searches for invisible SM Higgs decays  and in direct DM detection in astroparticle experiments, first ignoring the relic density constraint and then including it. A short conclusion in given at the end.

\section{The theoretical set--up}
\vspace*{-1mm}


The effective Higgs--portal scenario has been formulated for the spin--0,
spin--$\frac12$ as well as for the spin--1 DM cases \cite{Arcadi:2019lka}. Assuming CP--conserving interactions, the latter case is described by the following effective Lagrangian \cite{Mambrini:2011ik,Djouadi:2012zc} 
\begin{eqnarray} 
\label{Lag:DM}  
\Delta {\cal L}_V = \frac12  m_V^2 V_\mu V^\mu\! +\! \frac14
\lambda_{V}  (V_\mu V^\mu)^2\! +\! \frac14 \lambda_{HVV}  \Phi^\dagger \Phi
V_\mu V^\mu , 
\end{eqnarray} 
where $\Phi$ is the SM Higgs doublet field and $V_\mu$ the DM vector field; the $(V_\mu V^\mu)^2$ term is  not essential for our discussion and can be ignored. After electroweak symmetry--breaking, the original field $\Phi$ is decomposed as $(v + H)/ \sqrt{2}$ with $v=246$ GeV, inducing the trilinear interaction $\lambda_{HVV}$ between the physical $H$ state  and DM pairs and the DM mass term, $M_V^2 = m_V^2 + \frac{1}{4}\lambda_{HVV} v^2$. The vector Higgs--portal model formulated above is extremely simple and predictive, featuring only $M_V$ and $\lambda_{HVV}$ as free parameters. Together with the spin--0 and spin--$\frac12$ cases which share the same properties, it has been extensively studied theoretically and became a popular benchmark for experimental studies  \cite{Aad:2015pla,Chatrchyan:2014tja,Aaboud:2019rtt,Aaboud:2018sfi,Sirunyan:2018owy,Khachatryan:2016whc}. 

A requirement for the DM particle that is usually made is to lead to the correct cosmological relic density, as measured e.g.~by the Planck experiment \cite{Aghanim:2018eyx}. If the conventional freeze--out paradigm is assumed, it is determined by the thermal average $\langle \sigma v \rangle$ of the DM pair annihilation cross section, which encodes the information from particle physics. It receives contributions from annihilation processes into standard fermion and gauge (or Higgs) boson pairs whose rates depend on $M_V$ and $\lambda_{HVV}$. The experimentally favored value $\Omega_{\rm DM}h^2 \approx 0.12$ \cite{Aghanim:2018eyx} is achieved for $\langle \sigma v \rangle \simeq \mathcal{O}\left(10^{-26}\right){\mbox{cm}}^3 {\mbox{s}}^{-1}$. 

In order to be viable, a given assignment of the $[M_V,\lambda_{HVV}]$ set of input parameters should in principle give the favored value of  $\langle \sigma v \rangle $  above. Nevertheless, one can consider to relax the assumption
that the DM state leads to the measured relic density, e.g. by assuming that the vector singlet is not stable on cosmological scales and/or does not account for the entire DM in the Universe; see e.g. Ref.~\cite{Arcadi:2019lka} for a discussion.
 
In any case, more important is for the assignment of the $[M_V,\lambda_{HVV}]$ set to lead to a  DM scattering cross section over nucleons below the exclusion limit given  presently by the XENON1T experiment \cite{Aprile:2018dbl}. DM annihilation through CP--even boson exchange, being $s$--wave dominated, can be probed also through indirect detection  but the corresponding limits are subdominant compared to the direct detection ones. In addition, for light DM states with masses $M_V \leq \frac12 M_H$, searches for invisible decays of the SM--like $H$ boson at the LHC limit the branching fraction to BR$(H \rightarrow VV) \equiv {\rm BR} (H\to {\rm inv}) \lesssim 25\%$ \cite{Aaboud:2019rtt,Khachatryan:2016whc}.

\begin{figure}[!h]
\vspace*{-2mm}
    \centering
    \includegraphics[width=0.55\linewidth]{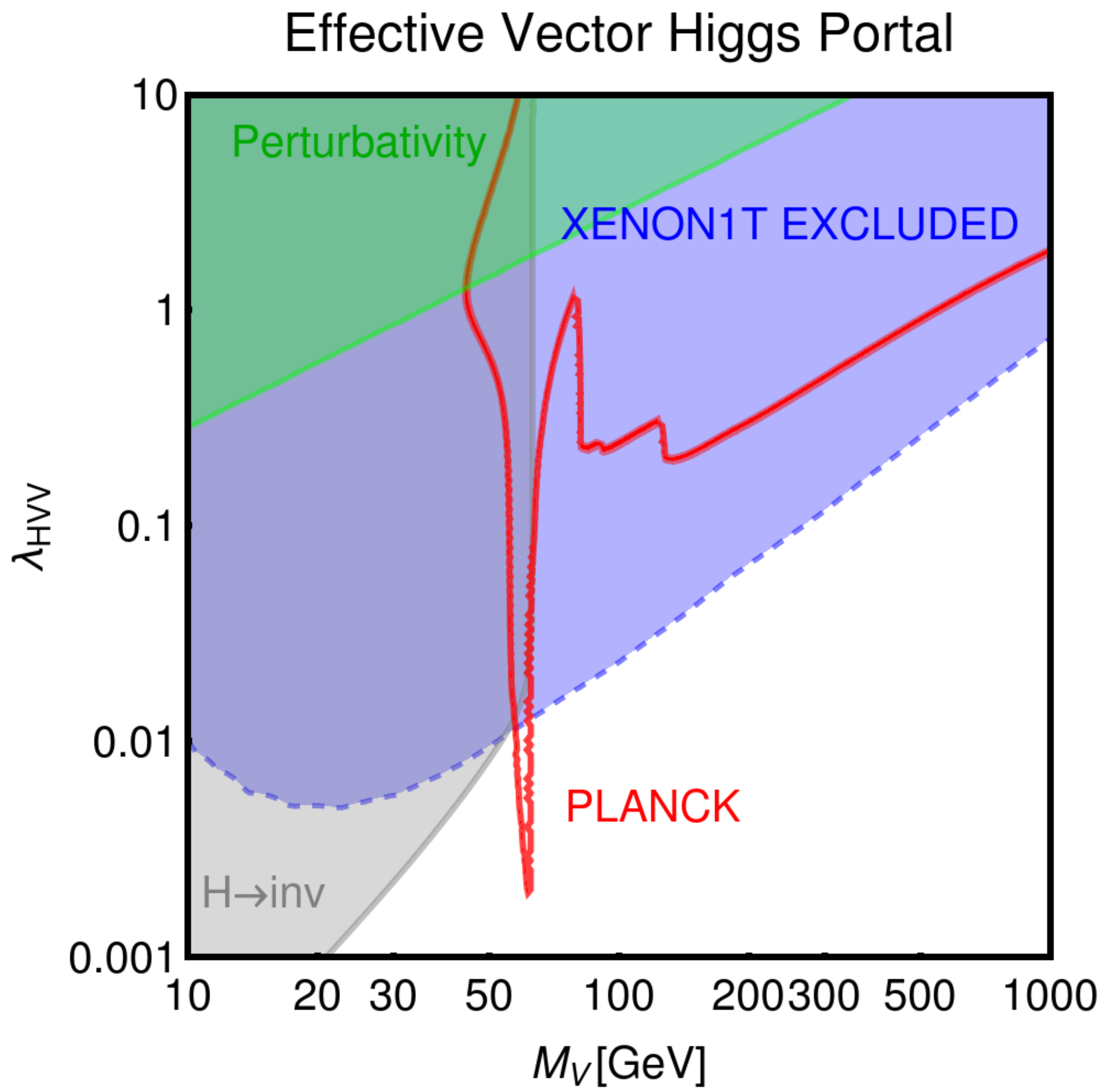}
\vspace*{-1mm}
    \caption{\footnotesize{Constraints on the effective vector DM Higgs--portal
 scenario in the  $[M_V,\lambda_{HVV}]$ plane. The red contour represents the area with the correct DM relic density, the blue region is excluded by current constraints from XENON1T, the gray area represents the one excluded by the invisible Higgs decays at the LHC,  while the green region does not fulfill the constraint eq.~(\ref{eq:PUbound}) from perturbative unitarity.}}
    \label{fig:EFT_vector}
\vspace*{-2mm}
\end{figure}

Besides these experimental limits, the vectorial Higgs--portal scenario is subject to theoretical constraints. For instance, when one considers bounds from perturbative unitarity on scattering processes such as $VV \rightarrow VV$ which proceeds via $H$--boson  exchange, one obtains the simple relation between the DM mass and coupling \cite{Lebedev:2011iq}
\begin{equation}
    M_V \geq \frac{\lambda_{HVV}v}{\sqrt{16\pi}} \label{eq:PUbound} ,  
\end{equation}
which forbids low DM masses $M_V$ at large $\lambda_{HVV}$ couplings. This is exemplified in Fig.~\ref{fig:EFT_vector} where, in the bidimensional $[M_V,\lambda_{HVV}]$ plane, one can see that the perturbative unitary bound eq.~(\ref{eq:PUbound}) excludes the green region corresponding to this possibility. In addition, when confronted with the requirement that the two parameters should be such that the value of the DM relic density $\Omega_V h^2$   reproduces the experimental one within $3\sigma$ (but as mentioned before, this requirement or assumption can be relaxed) as given by the thick red line of the figure, one also sees that the effective Higgs--portal with a vector DM does not provide a consistent description for masses $M_V \lesssim 50 \,\mbox{GeV}$ when $\lambda_{HVV} \gtrsim 1$. This constraint is however relaxed for smaller Higgs couplings.    

In Fig.~\ref{fig:EFT_vector}, we further include the constraints on the $[M_V,\lambda_{HVV}]$ parameter space from the most recent LHC measurements of the invisible Higgs branching ratio \cite{Aaboud:2019rtt,Khachatryan:2016whc} which exclude the grey area and from the direct detection of DM states by the XENON1T experiment \cite{Aprile:2018dbl} which excludes the blue area. As can be seen, the latter constraints are, in any case, much more stringent than the perturbative unitarity bound.


In order to assess the theoretical consistency of the effective Higgs--portal in the vector DM case, in particular for what concerns the interpretation of the searches for invisible  Higgs decays, we compare it in the following with one of its simplest and most economical ultraviolet complete realizations: the one in which the vector DM couplings are generated through the mixing of the SM--like Higgs boson with an additional scalar state.
We briefly review below the essential features of such a scenario, mostly following the conventions of Ref.~\cite{Falkowski:2015iwa}. 

The simplest potential accounting for mass mixing between the SM Higgs field $H$ and an additional scalar degree of freedom $S$ is given by
\begin{equation}
\label{eq:general_potential}
    V(h,s)=\frac{\lambda_H}{4}H^4+\frac{\lambda_{HS}}{4}H^2 S^2+\frac{\lambda_S}{4}S^4+\frac{1}{2}\mu_H^2 H^2+\frac{1}{2}\mu_S^2 S^2\, . 
\end{equation}
After electroweak  symmetry breaking, the fields $H$ and $S$ acquire vacuum expectation values $v$ and $\omega$  which, in terms of the above Lagrangian parameters, can be written as
\begin{equation}
    v^2 \equiv \frac{2 \lambda_{HS}\mu_S^2-4 \lambda_S \mu_H^2}{4\lambda_H \lambda_S-\lambda_{HS}^2},\,\,\,\,\omega^2 \equiv \frac{2 \lambda_{HS}\mu_H^2-4 \lambda_H \mu_S^2}{4\lambda_H \lambda_S-\lambda_{HS}^2} \, .
\end{equation}
In this setup, the $2\times 2$ mass matrix for the two scalar states is diagonalised via an orthogonal transformation of an angle $\theta$ with 
$\tan 2 \theta = {\lambda_{HS}v \omega}/{(\lambda_S \omega^2-\lambda_H v^2)}$
leading to the following masses for the two Higgs eigenstates (we use the abbreviation $c_\theta \equiv \cos\theta$ etc..) 
\begin{equation}
    M_{H_1,H_2}^2=\lambda_H v^2+\lambda_S \omega^2 \mp \frac{\lambda_S \omega^2-\lambda_H v^2}{c_{2 \theta}} \, ,  
\end{equation}
and we identify $H_1$ with the 125 GeV SM--like Higgs boson. In our analysis, we adopt the set $[M_{H_2},\sin\theta, \lambda_{HS}]$ as free parameters. The two quartic couplings $\lambda_H$ and $\lambda_S$ can be written as functions of the latter as (we use $\Delta M_{H}^2 = M_{H_2}^2 - M_{H_1}^2$)
\begin{align}
\lambda_H=\frac{M_{H_1}^2}{2v^2}+s^2_\theta \frac{\Delta M_{H}^2}{2v^2} \, , \ \  \lambda_S=\frac{2\lambda_{HS}^2}{s^2_{2\theta}}\frac{v^2}{\Delta M_{H}^2}\left(\frac{M_{H_2}^2}{\Delta M_{H}^2}-s^2_\theta\right) \, . 
\end{align}
 
Note that if the couplings of the scalar potential are assumed to be real and if ones requires that $\omega^2,v^2 >0$, one obtains  the two constraints $\lambda_H > \lambda_{HS}^2/(4 \lambda_S)$ and $\lambda_S>0$.  In terms of the set of free parameters adopted above, the couplings of the $H_1$ and $H_2$ states among themselves and with the SM fermions and gauge bosons are given by
\begin{align}
\label{eq:Lag-SM-tril}
    & \mathcal{L}_{\rm S}^{\rm SM}=(H_1 c_\theta + H_2 s_\theta)\left(2 M_W^2 W^{+}_{\mu}W^{-\,\mu}+M_Z^2 Z_\mu Z^\mu -m_f \bar f f\right)/v \, , \nonumber\\
    & \mathcal{L}_{\rm S}^{\rm tril}=-(\kappa_{111} H_1^3+\kappa_{112} H_1^2 H_2  s_\theta + \kappa_{221} H_1 H_2^2  cos_\theta + \kappa_{222} H_2^3)v/2 \, , 
\end{align}
with
\begin{align}
\label{eq:trilinear}
    \!&\! \kappa_{111}\!=\!\frac{M_{H_1}^2}{v^2 c_\theta}\left(c^4_\theta\!- \!s^2_\theta \frac{\lambda_{HS}v^2}{\Delta M_{H}^2} \right),
    \!&\! \kappa_{112}\!=\!\frac{2 M_{H_1}^2\!+\!M_{H_2}^2}{v^2}\left(c^2_\theta \!+\!\frac{\lambda_{HS}v^2}{\Delta M_{H}^2}\right) ,  \nonumber\\
    \!&\! \kappa_{221}\!=\!\frac{2 M_{H_2}^2\!+\!M_{H_1}^2}{v^2}\left(s^2_\theta \!-\!\frac{\lambda_{HS}v^2}{\Delta M_{H}^2}\right),
  \!& \!\kappa_{222}\!=\!\frac{M_{H_2}^2}{v^2 s_\theta}\left(s^4_\theta \!+\!c^2_\theta \frac{\lambda_{HS}v^2}{\Delta M_{H}^2} \right).
\end{align}


The vector DM Higgs--portal scenario can be introduced in the setup depicted above  by identifying the field $S$ and the vectorial DM candidate $V$ as, respectively, the Higgs and the gauge boson of a new, spontaneously broken, U(1) gauge symmetry. This scenario has been studied in detail in 
e.g.~Refs.~\cite{Lebedev:2011iq,Baek:2012se,Gross:2015cwa,Glaus:2019itb}. The newly introduced fields can be described by the following Lagrangian:
\begin{equation}
    \mathcal{L}_{\rm hidden}=-\frac{1}{4}V_{\mu \nu}V^{\mu \nu}+{\left(D^\mu S \right)}^{\dagger}\left(D_\mu S\right)-V(H,S) \, , 
\end{equation}
where $V_{\mu \nu}=\partial_\mu V_\nu-\partial_\nu V_\mu$ is the DM field strength and $D_\mu=\partial_\mu +i \tilde{g} V_\mu$ with $\tilde{g}$ being the new gauge coupling. The potential $V(H,S)$ has the same form as the one given in eq.~(\ref{eq:general_potential}). Upon the spontaneous breaking of the U(1) symmetry, the DM state acquires a mass $M_V= \frac12 \tilde{g} \omega $ and its interactions with the physical state $\rho$ of the original field $S\!=\!\frac{1}{\sqrt{2}}\left(\omega\!+\!\rho\right)$ are given by
\begin{eqnarray}
    \mathcal{L}_{\rm DM}=\frac14 {\tilde{g}^2}\omega\rho V_\mu V^\mu+\frac18 {\tilde{g}^2}\rho^2 V_\mu V^\mu=\frac12 {\tilde{g}}M_V \rho V_\mu V^\mu+\frac18 {\tilde{g}^2}\rho^2 V_\mu V^\mu \, .
\end{eqnarray}
After electroweak symmetry--breaking and the subsequent mass mixing between $h$ and $\rho$, the complete interaction Lagrangian, relevant for the DM phenomenology, is then
\begin{eqnarray}
   \mathcal{L}\!=\!\frac12 {\tilde{g}M_V} \left(H_2 c_\theta \!-\!H_1 s_\theta \right)V_\mu V^\mu \!+\! \frac18 {\tilde{g}^2}\left(H_1^2 s^2_\theta \!-\!2 H_1 H_2 s_\theta c_\theta \! +\! H_2^2 c^2_\theta\right)V_\mu V^\mu \!+\! 
\mathcal{L}_{\rm S}^{\rm SM}\!+\! \mathcal{L}_{\rm S}^{\rm tril}  \, ,   
\end{eqnarray}
where  $\mathcal{L}_{\rm S}^{\rm SM}$ and $\mathcal{L}_{\rm S}^{\rm tril}$ are given in 
eq.~(\ref{eq:Lag-SM-tril}). We can trade, as free parameters, the portal coupling $\lambda_{HS}$ with the dark gauge coupling $\tilde{g}$ using the  relation $\lambda_{HS}=\tilde{g}\sin 2 \theta \Delta M_{H}^2/(4 v M_V)$. 
The appropriate set of free parameters will be then composed by $\left[ M_V, \tilde{g}, \sin\theta ,M_{H_2} \right]$. 

The effective vector DM Higgs--portal can be obtained by taking the limits $\sin\theta \ll 1$ and $M_{H_2} \gg M_{H_1}$, provided that the theoretical consistency of this limit is verified.
First, the coupling $\tilde{g}$ has to remain perturbative,
$\tilde{g}^2/( 4\pi) \leq 1$. 
Second, the perturbative unitarity limit of the cross sections for processes of the type $H_i H_i \rightarrow H_j H_j$ which translates into  the limit 
$\lambda_i \leq \mathcal{O} \left({4 \pi}/{3}\right)$ \cite{Chen:2014ask} has to be accounted for. 

It is obvious from the limit on $\lambda_{HS}$ above  that perturbative unitarity constrains the hierarchy between $M_V$ and $M_{H_2}$. It might be hence not possible to have arbitrarily light DM and, at the same time, decouple the state $H_2$ from the phenomenology accessible to colliders. This
is the main concern raised by the LHC collaborations to which we turn now. 

\section{Comparing the EFT and UV--complete approaches}

In this section, we compare the EFT Higgs--portal and the dark U(1) model 
for vector--like DM states, from the perspective of the complementarity between DM direct detection in astroparticle physics  and invisible Higgs decay searches. We start with the case in which we are agnostic on the  production mechanism of DM in the early Universe. 

Focusing, for the time being, on the case $M_{H_2}\!> \! M_{H_1}$ with $H_1\equiv H$, one can make use of some analytical expressions. The invisible decay rate of the SM--like Higgs boson $H$ into DM $VV$ pairs, $\Gamma_{\rm inv}= \Gamma(H \to VV)$, can be written in the EFT and U(1) cases as
\begin{align}
&     \Gamma_{\rm inv}|_{\rm  EFT}=\frac{\lambda_{HVV}^2 v^2 M_H^3}{128 \pi M_V^4} \beta_{VH}  \, ,  \nonumber \\ 
&     \Gamma_{\rm inv}|_{\rm U(1)}=\frac{\tilde{g}^2 \sin^2 \theta}{32 \pi}\frac{M_{H_1}^3}{M_V^2} \beta_{VH_1} \, , 
\end{align} 
with the phase--space function $\beta_{VH}= \left(1-{4 M_V^2}/{M_H^2}+12 {M_V^4}/{M_H^4}\right){\left(1-{4 M_V^2}/{M_H^2}\right)}^{1/2}$.     
By solving the two equations above for respectively $\lambda_{HVV}^2$ and $\tilde{g}\sin^2 \theta$, we can express the vector DM spin--independent scattering cross section on protons $\sigma_{Vp}^{\rm SI}$ in terms of the invisible Higgs branching fraction BR$(H\to VV) \equiv  \Gamma(H\to VV)/ 
\Gamma_H^{\rm tot}$ with $\Gamma_H^{\rm tot}$ being the Higgs total decay width while $\mu_{Vp} = {M_V m_p}/{(M_V+m_p)}$ is the DM--proton reduced 
mass\footnote{In a recent paper \cite{Glaus:2019itb}, is has been shown that radiative corrections to the DM scattering cross section are relevant for $\tilde{g}\geq 1$, but realistic $H_1$ invisible branching fractions require $\tilde{g} <1$.},
\begin{align}
    & \sigma_{Vp}^{\rm SI}|_{\rm EFT}=8  \mu_{Vp}^2 \frac{M_V^2}{M_H^3} \frac{{\rm BR} \left(H\rightarrow VV\right) \Gamma_H^{\rm tot}}{\beta_{VH}} \frac{1}{M_H^4}\frac{m_p^2}{v^2}|f_p|^2 \, , \nonumber\\
    & \sigma_{Vp}^{\rm SI}|_{\rm U(1)}=8 \cos^2 \theta  \mu_{Vp}^2 \frac{M_V^2}{M_{H_1}^3} \frac{{\rm BR}\left(H_1\rightarrow VV\right) \Gamma_{H_1}^{\rm tot}}{\beta_{VH_1}} {\left(\frac{1}{M_{H_2}^2}-\frac{1}{M_{H_1}^2}\right)}^2 \frac{m_p^2}{v^2}|f_p|^2 \, . 
\end{align}

In the regime $M_{H_2}\!>\!M_{H_1}\!=\!M_H$, one has for the total Higgs widths, $\Gamma_{H_1}^{\rm tot}=\Gamma_{H}^{\rm tot}$. The predictions of the effective vector Higgs--portal and of the dark U(1) models will coincide, for a fixed value of the branching ratio BR$(H \rightarrow VV)={\rm BR}(H_1 \rightarrow VV)$, in the limit
\begin{equation}
\label{eq:r_analytic}
\cos^2 \theta \, M_{H}^4 \,  {\left({1}/{M_{H_2}^2}-{1}/{M_{H_1}^2}\right)}^2 \approx 1. 
\end{equation}

\begin{figure}[!h]
\vspace*{-3mm}
    \centering
    \subfloat{\includegraphics[width=0.45\linewidth]{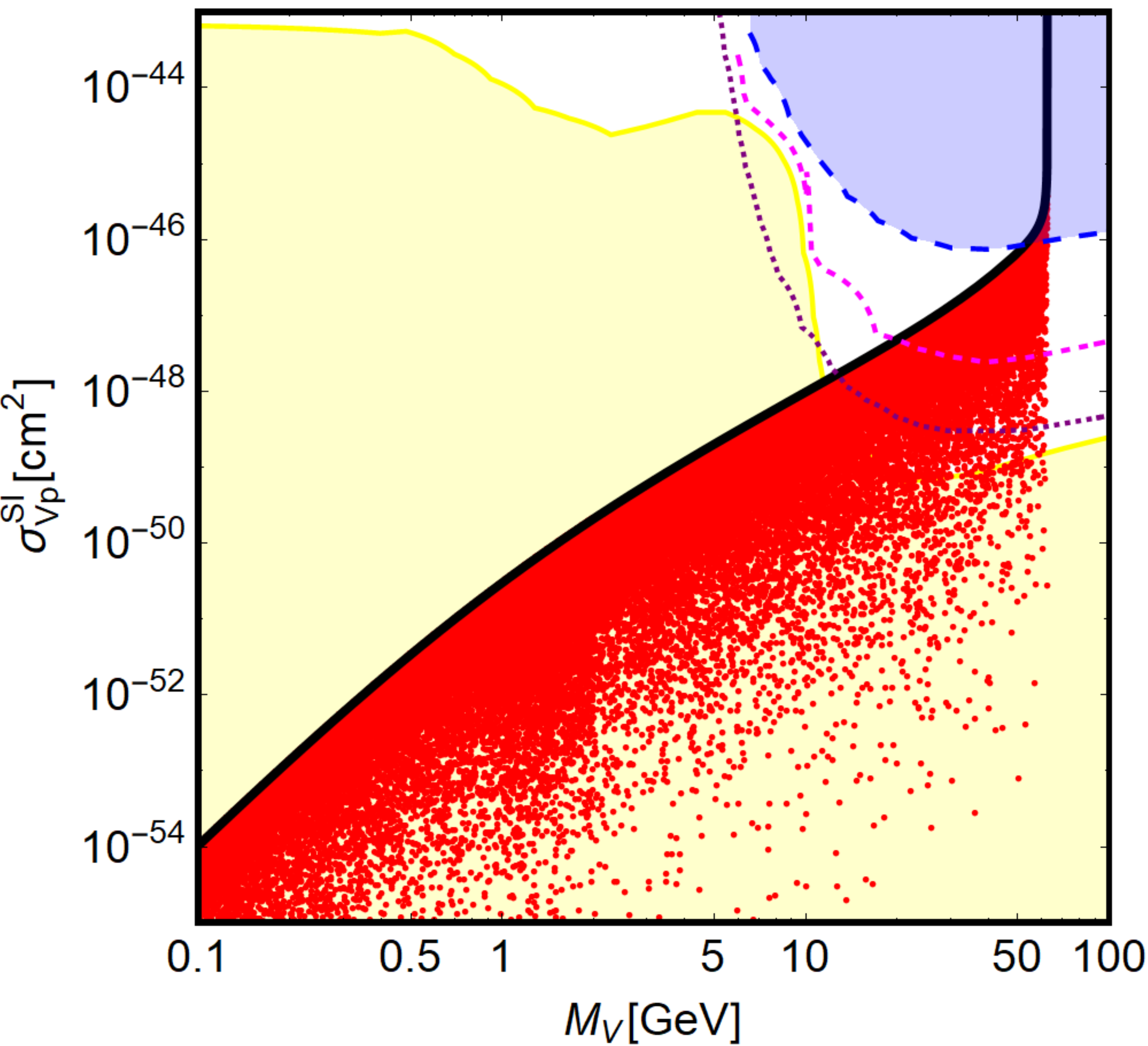}}~~
    \subfloat{\includegraphics[width=0.45\linewidth]{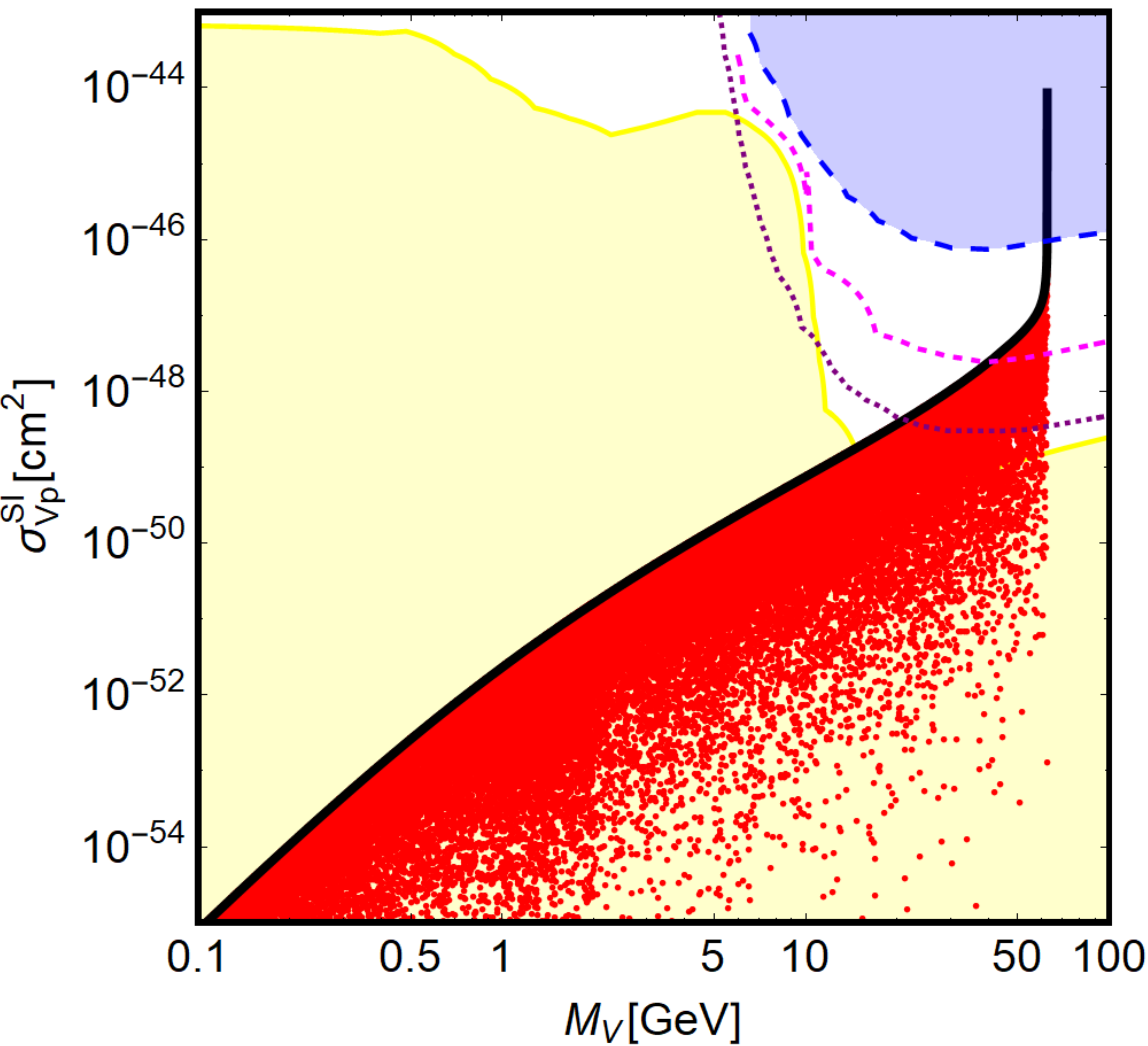}}
    \caption{\footnotesize{Values of the DM scattering cross section corresponding to BR$(H \rightarrow \mbox{inv})=0.25$ (left panel) and BR$(H \rightarrow \mbox{inv})=2.5 \times 10^{-2}$ (right panel), obtained by scanning over the dark U(1) model parameters space (red points). These are compared with the prediction of the effective Higgs--portal (solid black line), the exclusion from XENON1T (blue region) and the expected sensitivities of XENONnT (dashed magenta line) and DARWIN (dotted purple line). The yellow region corresponds to the so--called neutrino floor.}
}
    \label{fig:U1width}
\vspace*{-3mm}
\end{figure}

We have illustrated in Fig.~\ref{fig:U1width} the evolution of the ratio $r=\sigma_{Vp}^{\rm SI}|_{\rm U(1)}/\sigma_{Vp}^{\rm SI}|_{\rm EFT}$ in the plane $[M_V,\sigma_{Vp}^{\rm SI}]$, as considered by the LHC collaborations. The figure displays the outcome of a parameter scan conducted over the following ranges:
\begin{eqnarray}
    \sin\theta \in \left[10^{-3},0.3\right] \ \ , \
    M_V \in \left[10^{-2},62.5\right]\,\mbox{GeV} \ \ , \
    M_{H_2} \in \left[125.1,1000\right]\,\mbox{GeV} \, , 
\end{eqnarray} 
with the coupling $\tilde{g}$ fixed such that invisible branching ratios 
from the kinematically allowed decay $H_1 \to VV$ of  
BR$(H_1 \rightarrow \mbox{inv})=25\%$ and $2.5\%$ are obtained. While the first value corresponds to the present experimental sensitivity \cite{Aaboud:2019rtt,Khachatryan:2016whc}, $2.5\,\%$ represents the projected limit from the high--luminosity upgrade of the LHC \cite{Cepeda:2019klc}.

The obtained model points shown in red have been compared with the exclusion bound from XENON1T, depicted by the blue region above the dashed blue line, as well as with the projected sensitivities from the XENONnT \cite{Aprile:2015uzo} (a similar sensitivity is expected for the LZ experiment \cite{Szydagis:2016few}) and DARWIN experiments \cite{Aalbers:2016jon}, the dashed magenta and solid purple lines, respectively, and finally with the contour corresponding to invisible Higgs branching ratios of  BR$(H \rightarrow \mbox{inv})=25\%$ (left) and $2.5\%$ (right panel), shown by the thick solid black lines, obtained in the effective vector Higgs--portal scenario. For the sake of comparison, we have also highlighted in yellow, the region corresponding to the so--called neutrino floor, i.e. the region in which direct detection experiments become sensitive to the coherent scattering processes of SM neutrinos over nucleons mediated by the SM $Z$--boson \cite{Billard:2013qya,Boehm:2018sux}.

In agreement with the findings of e.g. Ref.~\cite{Baek:2014jga}, we see that the DM scattering cross section correlated to a fixed value of the invisible Higgs branching fraction, in this case the present (left) and future (right) experimental sensitivities, can be orders of magnitude below the one expected in the effective Higgs--portal scenario. As already seen, this is due to the destructive interference effect of the additional scalar $H_2$  in the DM scattering cross section, which becomes negligible only if the mass $M_{H_2}$ is sufficiently high. The $\cos^2 \theta$ factor in eq.~(\ref{eq:r_analytic}) does not play a prominent role since its value is always close to unity, as a result of the LHC constraints from the Higgs signal strengths. 

Nevertheless, as already anticipated from eq.~(\ref{eq:r_analytic}), Fig.~\ref{fig:U1width} shows that the effective vector Higgs--portal can indeed represent a limiting case of the more realist dark U(1) model. One should, however, verify that this limit can be achieved consistently with the already mentioned unitarity and perturbativity constraints. The latter can be re--expressed as upped bounds on the invisible branching fraction of the $H_1$ state, i.e. only the values indicated below can be regarded as theoretically consistent:
\begin{align}
    & \lambda_{HS} \leq \frac{4 \pi}{3} \Longrightarrow {\rm BR}(H_1 \rightarrow VV) \lesssim 0.25 {\left(\frac{3\,\mbox{TeV}}{M_{H_2}}\right)}^4 \, ,\nonumber\\
    & \lambda_{S}\ ~ \leq \frac{4 \pi}{3} \Longrightarrow {\rm BR}(H_1 \rightarrow VV) \lesssim 0.35 {\left(\frac{\sin \theta}{0.1}\right)}^2 {\left(\frac{3\,\mbox{TeV}}{M_{H_2}}\right)}^2 \, .
\label{eq:lambda_HSS}
\end{align}

\begin{figure}[!h]
\vspace*{-3mm}
    \centering
    \includegraphics[width=0.55\linewidth]{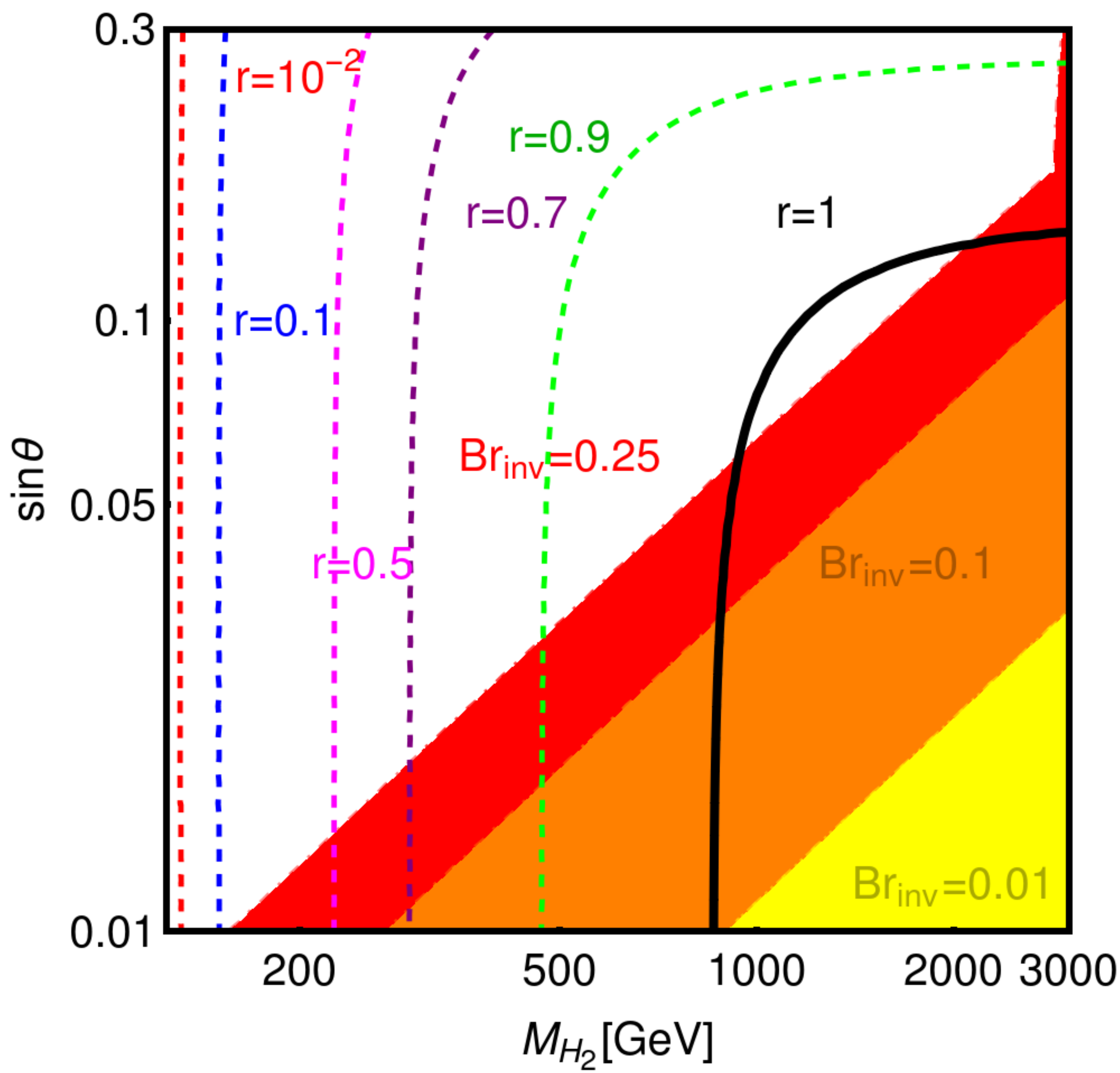}
    \caption{\footnotesize{Contours in the bidimensional plane $[M_{H_2},\sin\theta]$ of the ratio $r$ of the DM scattering cross section of vector DM for the dark U(1) over the same cross section but computed for the effective vector Higgs--portal. The red, orange, yellow regions are theoretically inconsistent because of perturbative unitarity for values of the invisible Higgs branching ratio Br$_{\rm inv} \equiv {\rm BR}(H_1 \to VV)= 0.25,0.1,0.01$,  respectively.}}
    \label{fig:figBr}
\vspace*{-1mm}
\end{figure}

The consistency of the EFT limit is more explicitly illustrated in Fig.~\ref{fig:figBr} which shows isocontours of the ratio $r=\sigma^{\rm SI}_{Vp}|_{\rm U(1)}/\sigma_{Vp}^{\rm SI}|_{\rm EFT}$ in the plane $[M_{H_2}, \sin\theta]$. In red/orange/yellow are shown, respectively, the areas excluded by the perturbativity bounds on $\lambda_{HS}$ and $\lambda_{S}$, eq.~(\ref{eq:lambda_HSS}) for three values of the invisible Higgs branching ratio, namely  BR$(H \! \to \! \mbox{inv} )\!= \!25\%$, corresponding to the current LHC experimental sensitivity, BR$(H \! \to \!  \mbox{inv} )\!=\! 10\%$ and $1\%$, which would correspond to the sensitivities to be obtained at, respectively, the next phase of the LHC and a future high--energy $pp$ or an intermediate energy $e^+e^-$ collider \cite{Arcadi:2019lka}. We have included the restriction $\sin\theta <0.3$ which should lead only to acceptable deviations of the signal strengths for the various Higgs production and decay rates (with Higgs couplings $g_{HXX}^2 \propto \sin^2\theta \lesssim 10\%)$ , see e.g. Refs.~\cite{Khachatryan:2016vau,Sirunyan:2018koj,Aad:2019mbh}, as well as to avoid bounds from direct LHC searches of $H_2$, see e.g. Refs.~\cite{Aaboud:2018bun,Sirunyan:2018qlb,Sirunyan:2019pqw,Aad:2019uzh,Sirunyan:2018two} with the exception of the ones into $H_2 \rightarrow ZZ$ which can still constrain $130 \lesssim M_{H_2} \lesssim 200\,\mbox{GeV}$ for low decay branching fraction of $H_2$ into DM \cite{Huitu:2018gbc}. The latter constraintq have, however, a negligible impact on our results.
 
As can be seen, the strongest deviations occur for $M_{H_2} \lesssim 200\,\mbox{GeV}$. One can see that a theoretically consistent recovery of the EFT limit, i.e. with perturbative unitarity, giving $r\approx 1$ for the three values of the invisible Higgs branching ratio, is possible. However, for the present experimental sensitivity BR$(H \! \to \! \mbox{inv} )\!= \!0.25$, this occurs only for the values 1~TeV~$\lesssim M_{H_2} \lesssim 2\,\mbox{TeV}$ and $0.06 \lesssim \sin\theta \lesssim 0.12$. Nevertheless, the situation rapidly improves with the smaller invisible branching ratios that are expected to be probed at the next phases of the LHC and at future colliders. 

The discussion and simple expressions presented above were under the assumption $M_{H_2}>M_{H_1}$ so that the only modification with respect to the SM Higgs boson case, is the presence of the additional decay channel into DM pairs, $H\to VV$. In the dark U(1) model, one might consider the case that the additional scalar $H_2$ is light enough, i.e. $M_{H_2} \leq \frac12 M_{H_1}$, so that also the process $H_1 \rightarrow H_2 H_2$, whose rate is given by
\begin{equation}
    \Gamma (H_1 \rightarrow H_2 H_2)=\frac{\cos^2 \theta \kappa_{122}^2 v^2}{32\pi^2 M_{H_2}}\sqrt{1-{4 M_{H_2}^2}/{M_{H_1}^2}},
\end{equation}
contributes to the invisible branching fraction of the SM--like Higgs boson. Here, we briefly comment on the possibility that an eventual collider signal associated with the Higgs invisible branching ratio would actually be produced by the combination of the $H_1 \!\rightarrow \! VV$ and $H_1 \!  \rightarrow \! H_2 H_2$ decay processes and mimic the result given by the experimental collaborations, i.e.~lead to a ratio $r\approx 1$. To see the implications of such a hypothesis, we have performed a parameter scan similar to the one before but adopting the mass range $M_{H_2} \in \left[10,62.5\right]\, \mbox{GeV}$ for the additional $H_2$ state and varying the gauge coupling $\tilde{g}$  such that the relation $0.01 \leq {\rm BR}(H_1 \rightarrow VV)+{\rm BR} (H_1 \rightarrow H_2 H_2)\leq 0.25$ holds. 

\begin{figure}[!h]
\vspace*{-1mm}
    \centering
    \includegraphics[width=0.55\linewidth]{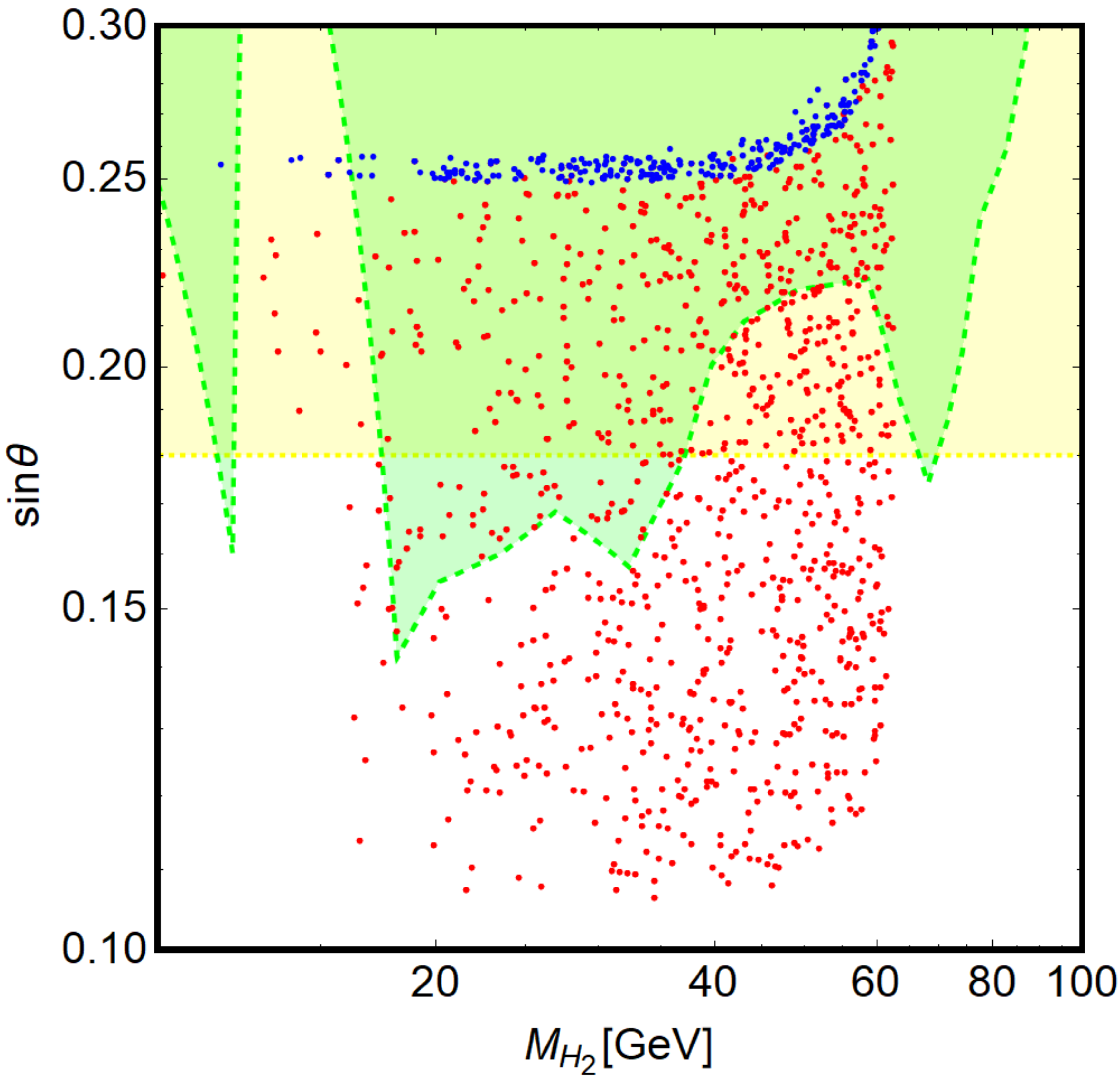}
    \caption{\footnotesize{Model points in the $[M_{H_2},\sin\theta]$ bidimensional plane in which $M_{H_2}\leq \frac12 M_{H_1}$ with $0.9 \leq r \leq 1.1$. The blue points feature BR$(H_1 \rightarrow \mbox{inv})=0.25$ and the red ones  $0.01 \leq {\rm BR}(H_1 \rightarrow \mbox{inv})<0.25$. The green area is excluded by complementary constraints on $H_2$, as given in Ref.~\cite{Falkowski:2015iwa}}. The yellow area is the one that can be excluded by future   measurements of the Higgs signal strengths at HL--LHC.}
    \label{fig:U1light}
\vspace*{-1mm}
\end{figure}

The results are shown in Fig.~\ref{fig:U1light} where the model points for which $0.9 \leq r \leq 1.1$ in the $[M_{H_2},\sin\theta]$ bidimensional plane are displayed. We have marked in blue the points for which ${\rm BR}(H_1 \rightarrow \mbox{inv}) \equiv {\rm BR}(H_1 \rightarrow VV)+{\rm BR}(H_1 \rightarrow H_2 H_2)=0.25$. As can be seen, in order to have a value $r\approx 1$ and, at the same time, an invisible branching fraction of at least $1\,\%$, relatively high values of $\sin\theta$, namely $\gtrsim 0.1$, are required. One should then verify the compatibility of this result with eventual complementary constraints. Shown in green is the excluded region from different constraints, including direct searches of $H_2$, as determined in Ref.~\cite{Falkowski:2015iwa}. As can be seen, the case BR$(H_1 \rightarrow \mbox{inv})=25\%$ is ruled out except in a small range. Lower invisible branching fractions, expected to be probed in the near future, would instead be still viable. Together with the one of invisible Higgs decays, the experimental sensitivity on the Higgs signal strengths will improve as well in the future and we have thus included in Fig.~\ref{fig:U1light} the region (in yellow) which can be excluded by the bound $\sin\theta < 0.18$, expected to be reached at the high--luminosity LHC \cite{Cepeda:2019klc}.

Hence, the analysis above shows that, indeed, the simplest UV--complete model for a vector DM state with a Higgs--portal can be described by a simple EFT approach in which collider and astroparticle results can be consistently compared.  We expect this conclusion to stay valid in other more complex 
UV--complete extensions in which, for example, the gauge group is larger and/or  the symmetry--breaking mechanism of the hidden sector is
more complicated. In these more involved cases too, one should be able to 
consistently decouple the extra degrees of freedom in order to describe the model in the EFT approach for DM masses within the reach of colliders.

Finally, let us comment for completeness on the possibility of imposing the requirement that the DM relic density is compatible with the measured one.  For this purpose, considering  the case $M_{H_2}\! > \! M_{H_1}$, we have performed the following parameter scan
\begin{align}
\label{eq:scan_range}
    M_V \in \left[1,10^3\right] \mbox{GeV} \, , \
    M_{H_2} \in \left[125.1,1000\right]\mbox{GeV} \, , \
    \sin\theta \in \left[10^{-3},0.3\right] \, , \
    \tilde{g} \in \left[10^{-3},10\right] \, , 
\end{align}
retaining only the model points which lead to an  $\Omega_{DM}h^2$ value, as calculated by the program micrOMEGAs 5.0 \cite{Belanger:2018mqt}, within a $3\sigma$ interval around the experimentally favored value. Besides the present XENON1T bound on the DM scattering cross section over protons, we also considered, for completeness, the FERMI bounds from DM indirect detection summarized in Ref.~\cite{Clark:2017fum}. 


\begin{figure}[!h]
\vspace*{-3mm}
    \centering
    \subfloat{\includegraphics[width=0.45\linewidth]{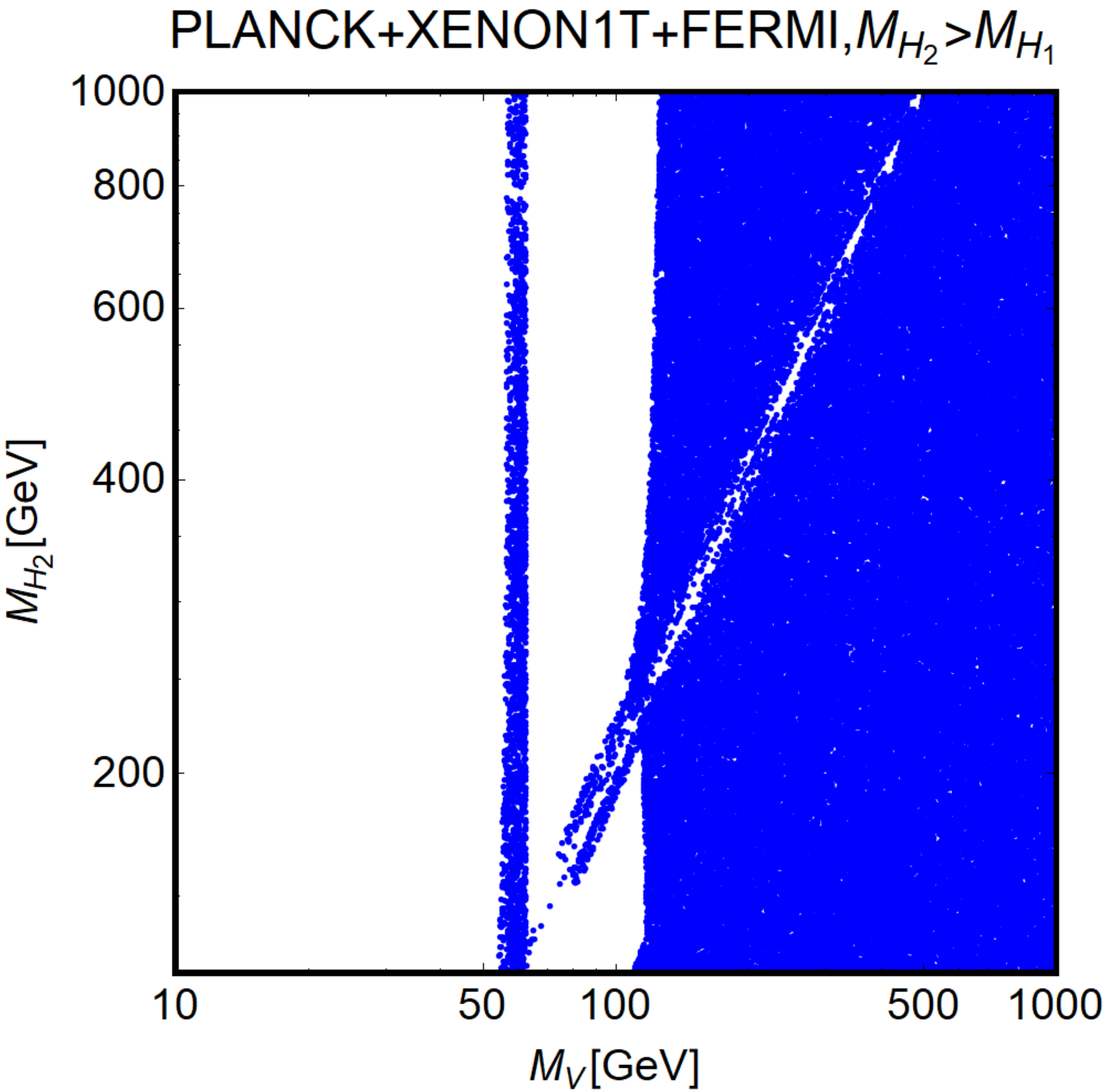}}~~~
    \subfloat{\includegraphics[width=0.45\linewidth]{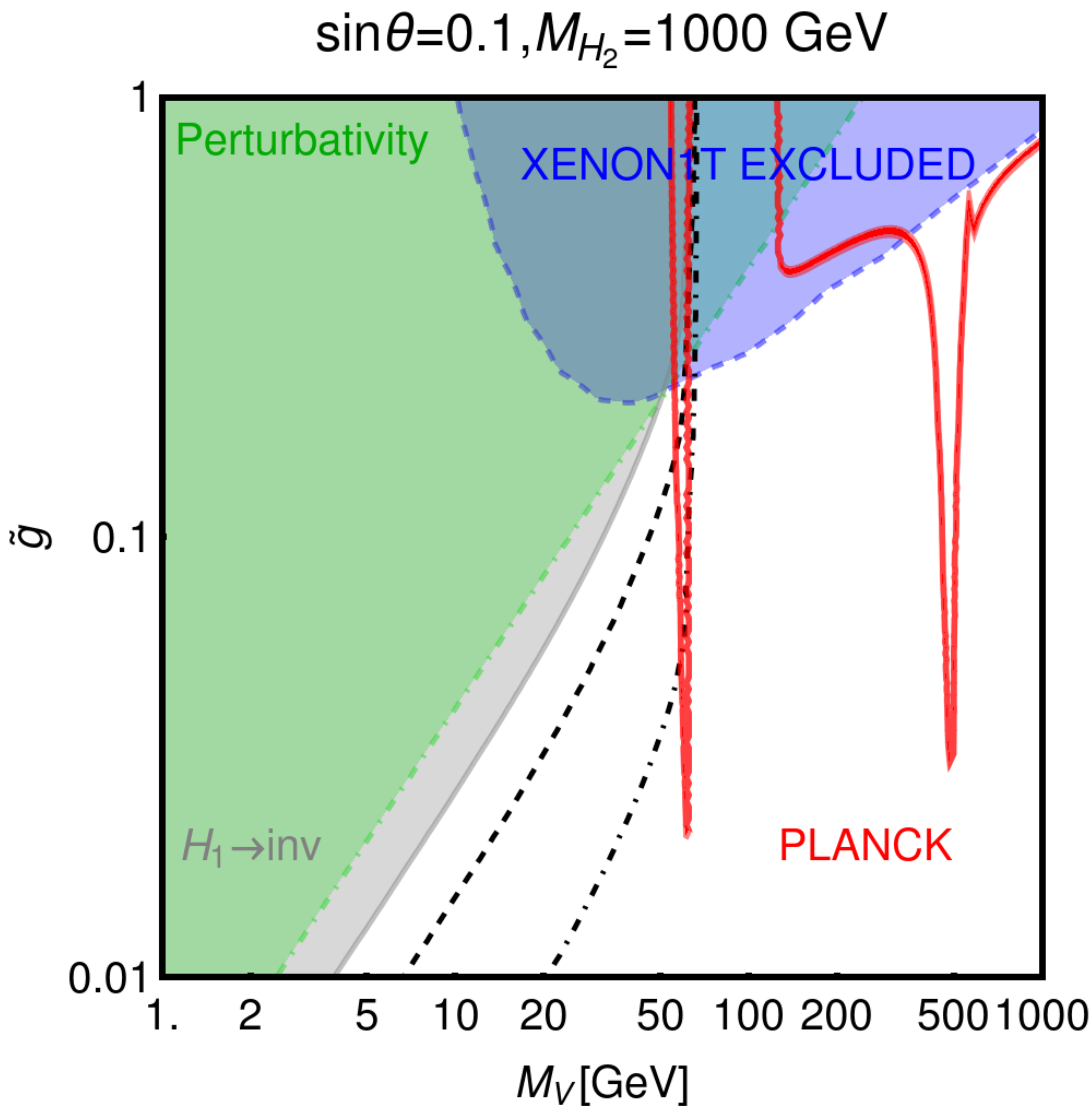}}
 \vspace*{-2mm}
    \caption{\footnotesize{Left panel: model points of the dark U(1) model in the $[M_V,M_{H_2}]$ plane complying with the constraints from DM direct and indirect detection and relic density and with the further requirement of $M_{H_2}>M_{H_1}$}. Right panel:
summary of constraints for three benchmark points of the model
with invisible Higgs branching fraction of 25\% (solid), $10\,\%$ (dashed) and $2.5\,\%$ (dot--dashed line); the regions with the correct relic density and
the one excluded  by XENON1T are also shown and compared to the green region excluded by the constraints from perturbative unitarity.}
    \label{fig:my_scan}
\vspace*{-3mm}
\end{figure}

As evidenced by the left--hand side of Fig.~\ref{fig:my_scan}, where the 
results of our scan are shown in the plane $[M_V, M_{H_2}]$, one can have viable (blue) points only for DM masses at most as low as $\frac12 M_{H_1}$, hence marginally lying within the kinematical reach for invisible Higgs decays. 

To better illustrate this outcome, we describe in more detail the combined theoretical and phenomenological constraints, including the current limit of 
BR$(H\to {\rm inv})=0.25$ (black curve) as well the projected limits (the dashed and dot-dashed black lines) corresponding to invisible Higgs branching fractions of $10\,\%$ and $2.5\,\%$, in the right--hand side of Fig.~\ref{fig:my_scan}, by considering a specific assignment of the set  $[M_{H_2},\sin\theta]$, namely $[1\,\mbox{TeV}, 0.1]$,  and varying instead the parameters $M_V$ and $\tilde{g}$. Included also 
are the bound from perturbative unitarity (the green region) and the one excluded by XENON1T (the blue region). 

One can see that the theoretical bounds become competitive with the ones from the SM--like Higgs invisible branching fraction. The DM relic density can be achieved, while being compatible with the other constraints, only in the pole $M_V=\frac12 M_{H_{1,2}}$ regions and, for DM masses $M_{V}>M_{H_2}$, when the $VV \to H_2 H_2$ annihilation channel becomes kinematically accessible. This means that for the considered value of $M_{H_2}$, most of the viable DM regions lie outside the reach of conventional searches of invisible Higgs decay process. To test effectively the case under consideration, one should rather consider DM pair production through exchange of off--shell Higgs bosons that lead to small rates \cite{Baglio:2015wcg}.

\section{Conclusions}

Searches for the particles that account for the DM in the Universe are gaining increasing relevance in the LHC physics program. Among these,  the determination of the invisible decays of the SM--like Higgs boson play a prominent role. The interpretations of the results are usually made using as a benchmark, an effective approach of the Higgs--portal scenario  which is rather simple and predictive as it involves only two relevant parameters: the mass of the DM and its coupling to the Higgs boson. Nevertheless, being not complete in the UV regime, this EFT approach might not provide a consistent interpretation of the experimental outcome, especially when it is compared to the direct detection in astrophysical experiments. This was thought to be particularly true in the case of a spin--1 DM state, a scenario that was removed from recent LHC interpretations of invisible Higgs decays searches.  

In this paper, we have reanalyzed the possibility that a Higgs--portal with a vectorial DM state could represent a consistent EFT limit of its simplest UV completion, dubbed dark U(1) model, for LHC searches of invisible Higgs decays. Considering the present experimental sensitivity on these decays, we show that it is indeed the case in a limited but non negligible region of the parameter spaces of the UV--complete model. The situation is, however, expected to considerably improve with the increase of the experimental sensitivity on the invisible Higgs decay branching ratio to be measured at the next phases of the LHC or at future colliders. In this case, the EFT approach could represent a viable limit of the renormalisable model in large regions of its parameter space. 

The previous statement is in general valid only if the DM state of the U(1) model is not required to generate the observed cosmological relic density. Assuming  conventional thermal production, a viable relic density for a light dark gauge boson with a mass $M_V \lesssim \frac12 M_H \approx 62.5\,\mbox{GeV}$ cannot be achieved with the second CP--even Higgs boson that is present in the model, decoupled from the low energy phenomenology. In this case, the interpretation of the searches of invisible SM--like Higgs decays would not be valid in the context of the EFT approach and one needs to consider the full model. Nevertheless, one could relax the assumption that the DM leads to the measured 
relic density by considering alternatives to the conventional thermal paradigm,   like for example  non--thermal production mechanisms and/or modified cosmologies. A further interesting alternative would be to consider more complicated scenarios such as larger dark symmetry groups, even for thermal WIMPs. These aspects will be discussed in a forthcoming paper.\bigskip


\noindent {\bf Acknowledgements:}  We thank Ketevi Assamagan, Martti Raidal and Rui Santos for discussions.  We warmly thank Oleg Lebedev for his valuable comments on the manuscript. This work is supported in part by the ERC Mobilitas Plus grant MOBTT86.

\setlength{\parskip}{0.12cm}

\bibliography{MajDM}
\bibliographystyle{JHEPfixed}

\end{document}